\documentstyle[aps,prl,multicol,epsf,amssymb]{revtex}
\input epsf

\def\d {{\rm d}}

\def\dpl{\epsilon}
\def\ddpl{\dot\dpl}

\def\sigp{\zeta}

\begin{document}

\tightenlines
\draft

\title {
The Origin of a Repose Angle:\\
Kinetics of Rearrangement for Granular Materials
}

\author{Ana\"el Lema\^{\i}tre$^{(1,2)}$}
\address{
$^{(1)}$Department of physics, University of California, Santa Barbara, California 93106, U.S.A.\\
$^{(2)}$CEA --- Service de Physique de l'\'Etat Condens\'e,
Centre d'\'Etudes de Saclay, 91191 Gif-sur-Yvette, France
}
\date{\today}
\maketitle
\begin{abstract}
A micro-structural theory of dense granular materials is presented,
based on two main ideas.
Firstly, that macroscopic shear results from activated local rearrangements 
at a mesoscopic scale.
Secondly, that the update frequency of microscopic processes is determined by granular
temperature.
In a shear cell,
the resulting constitutive equations account for Bagnold's scaling and for the existence
of a Coulomb criterion of yield.
In the case of a granular flow down an inclined
plane, they accounts for the rheology observed in recent experiments~\cite{pouliquen99a} 
and for temperature and velocity profiles measured numerically.~\cite{ertas00,silbert01}
Finally, it is shown that for an angle $\theta$ of the plane, 
the system jams below some critical height $H_{\rm stop}(\theta)$,
and the critical curve obtained fits remarkably well experimental data.
\end{abstract}

\begin{multicols}{2}
An upsurge of interest for granular materials has recently stirred
the physical literature.~\cite{rajchenbach00,jaeger96}
The problem is indeed challenging.
The situation is paradoxical. Those materials lie at our doorstep,
are ubiquitous in everyday' life,
their understanding is of an extreme interest for numberless practical reasons
ranging from earthquakes or landslides to industrial processes.
Yet, there is no satisfactory explanation for the obvious
fact that heaps have a slope\dots
Recent advances have been made in experimental and numerical studies.
In particular, the flow of a granular layer down an inclined plane is a laboratory model
for many realistic situations.~\cite{azanza99,pouliquen99a,daerr99,silbert01}
In this set-up, evidence has been given for
the existence of a critical curve $H_{\rm stop}(\theta)$ relating the angle of the slope,
$\theta$, to the thickness of a flowing layer below which the system jams.
This relation refines the well-known Coulomb criterion of yield.
It is accompanied by Bagnold's scaling,
in the bulk of a dense flow, which relates the shear stress $\sigma$
to the strain-rate $\ddpl$ by, $\sigma\propto\ddpl^2$.~\cite{bagnold54,silbert01}
Those experimental findings remain unexplained and add even more constraints 
to the challenge faced by the theorist.
Kinetic theory~\cite{savage79,jenkins83}
accounts for the rheology of dilute systems~\cite{bocquet00},
but fails to explain jamming and the rheology of dense systems.

This work focuses on structural rearrangements.
It draws on the so-called shear transformation zone (STZ)
theory,~\cite{falk98,falk00}
recently introduced to account for the behavior of viscoplastic solids.
The profound difference between those two systems, at the microscopic level,
is of particular interest:
it has been pointed out recently that some unifying concept might be at work,
in various systems: granular materials, glasses, foams,\dots~\cite{ohern01}
Hence, more than an isolated model of granular flows, this work attempts to bridge
the gap between two of those apparently different systems,
thus showing that structural rearrangement is a key to our understanding of the rheology
of dense systems on very general grounds.

Firstly, I will present a basic mechanism for rearrangement, 
extracted from~\cite{falk98,falk00}.
Then, after reviewing the basic scaling properties of the $N$-body problem,
I will show how this basic mechanism adapts to the case of granular materials.

A shear transformation zone (STZ) is defined as a locus within the material where 
a rearrangement is made possible by the local configuration of the contact network.~\cite{falk98,falk00}
An important remark that rests at the root of STZ theory is that once some microscopic shear 
has occurred somewhere in the material, the system cannot shear further at this point, 
and in this direction (although it may shear backward).
This leads to the identification of pairs of types of arrangements that are
transformed into one another by a local shear.
A local ``symmetry'' is induced by shearing,
and the local state of the system is determined by
the population of arrangements susceptible to shear in a given direction.
To simplify the picture, a single pair of orientations is considered, aligned along 
the principal axes of the stress tensor.
Such states are schematically represented below:
\begin{center}
\begin{picture}(100,30)(0,0)
\put(50,30){\makebox(0,0){\large$R_+$}}
\put(50,0){\makebox(0,0){\large$R_-$}}
\put(0,0){\makebox(100,30){
\epsfxsize=100\unitlength
\epsffile{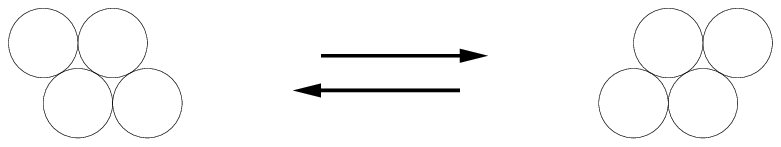}
}}
\end{picture}
\end{center}
\label{fig:stz}
although an actual STZ might involve more than four grains (say of order 10).

The populations of STZ's are denoted $n_+$ and $n_-$. And $R_+$ (resp. $R_-$)
denotes the rate at which a $+\to-$ (resp. $-\to+$) transformation occurs.
Those transitions are activated by the free-volume and force fluctuations.
Since I do not consider variations of volume in this work,
$R_\pm$ will be written as functions of the stress tensor only. 
Expressions for those rates will be given further.
The macroscopic shear is seen as the sum of all local rearrangements,
\begin{equation}
\label{eqn:stz:ddpl}
\ddpl = {\cal A}_0\,(R_+n_+-R_-n_-)
\quad,
\end{equation}
where $\ddpl$ is the off-diagonal component of the rate-of-deformation tensor,
and where ${\cal A}_0$ is some constant.

Equations of motion for the populations $n_\pm$ are written to close the system.
They are of the form,~\cite{falk98,falk00}
\begin{equation}
\label{eqn:stz:npm}
\dot n_\pm = R_\mp n_\mp - R_\pm n_\pm +\omega\,({\cal A}_{c} - {\cal A}_{a}\,n_\pm)
\end{equation}
The first two terms on the rhs account for the ``internal'' dynamics of STZ's, while the last term
introduces a coupling of the local arrangements with the mean flow.
This term can be understood in the following way:
From a macroscopic standpoint, the flow constantly stirs the grains, thus 
creating and destroying local configurations. 
Variable $\omega$ is the rate at which the flow induces new configurations.
In a mean-field situation, it can be evaluated as the overall work $\sigma\ddpl$
produced by the shear stress $\sigma$, and normalized by some typical force
of the contact network.
I will argue in the following that for granular materials, 
this typical force is measured by the pressure $P$ itself,
so that $\omega=\omega^{0}=\sigma\ddpl/P$ if space is not taken into account.
When spatial extension is considered this term introduces a spatial coupling:
since an STZ is a configuration of several grains at a mesoscopic scale,
the flow at a point $\vec r$ contributes to the creation/destruction of nearby STZ's.
In general, the term $\omega$ will be written,
\begin{equation}
\label{eqn:om}
\omega(\vec r) = {1\over2\ell}\,\int \omega^0(\vec r') e^{-{|\vec r-\vec r'|\over \ell}} \d\vec r'
\quad,
\end{equation}
where a length-scale $\ell$ has been introduced, which characterizes the extension of an STZ.

The remaining task consists in evaluating the rates $R_\pm$ and arguing about
the above-mentioned expression for $\omega^0$. But beforehand, it is necessary 
to put into perspective the essential features and scaling property of granular materials.


It has been observed recently that Bagnold's scaling is an immediate consequence of 
the equations of the $N$-body problem, so long as no time-scale is imposed to the system
by coupling it {\it e.g.} to a pressure bath.~\cite{bocquet00} 
It is however important to understand this scaling in greater details so as to 
write macroscopic equations that agree with the underlying physics.

Consider a system of $N$ grains in contact submitted to a set of forces $F^{\rm c}$ 
at each contact point, and to forces $F^{\rm c'}$ at contact points with the boundary.
The history of the system is given by the locations $\vec r_i(t)$ of the center of mass
of each grain, their rotations, 
the sets $c(t)$ and $c'(t)$ of the contact points, and the forces as 
function of time.  The interaction between the grains is a pure hard-core repulsion;
if there is friction, the yield criterion at each contact point involves a ratio of forces.
Equations of motion induce no time-scale, no scale for the forces network:
the system is unchanged if all forces are rescaled as,
$F^{\rm c, c'}\to F^{\rm c, c'}/F_0$ and if time scales as $t\to t\sqrt{F_0}$.
Rescaling the forces, leaves the trajectory of the system unchanged in the phase space!
Only the time-coordinate along the trajectory is modified.

This scaling invariance of the $N$-body problem lies at the root of Bagnold's scaling.
It is noteworthy that kinetic theory correctly incorporates this scaling invariance
(although it does not account for dense rheology where Bagnold's scaling is observed).
This is performed {\it via} the introduction of a granular temperature $T$ which,
as opposed to thermodynamic temperature, is a dynamical quantity,
and which determines the collision frequency, ${\cal F}=\sqrt{T}$.
Granular temperature is usually defined as the specific kinetic energy.
However in a dense material,
various contributions can be identified to kinetic energy: temperature obviously differs from
velocity fluctuations associated to the rearrangements themselves.
In this work, granular temperature is seen as associated to tiny motion of the grains in
the surrounding cage made of their neighbors, in the spirit of
Haff's~\cite{haff83,haff86} picture of kinetic theory.
By definition, granular temperature determines the smallest time-scale in the material
which is also the frequency of all microscopic processes.
From a phase space perspective, $1/\sqrt{T}$ determines the time-scale at which the system
evolves along a trajectory.

The equation of motion for temperature results from an energy balance between
the energy produced by the flow and energy dissipated by collisions:
\begin{equation}
\label{eqn:T}
\dot T = \sigma\ddpl - \alpha T\,\sqrt{T}
\end{equation}
The second term in the rhs accounts for collision-mediated energy dissipation:
collisions occur at the frequency $\sqrt{T}$, and each collision dissipate an energy 
proportional to $T$ itself.
Parameter $\alpha$ is related to the restitution coefficient of the grains.
It's straightforward to check that this equation in left invariant under the rescaling
$\sigma\to \sigma/F_0$, $t\to t\sqrt{F_0}$.
This energy balance, added to the fact that granular temperature controls
the smallest time-scale, {\it i.e.} the update frequency of microscopic processes,
is a translation, at the macroscopic level, of the scaling invariance of the $N$-body problem.
In dense flow, the fact that temperature determines the update frequency 
of activated processes, leads to the emergence of Bagnold's scaling.

It is important to make the difference between granular temperature 
and thermodynamic temperature: granular temperature does not measure a volume in the phase
space, henceforth a Boltzmann weight for activated processes.
The fact that thermodynamic temperature measures a volume in the phase space does not 
results immediately from its definition as kinetic energy, by from the coupling 
to a thermal bath, which blends nearby trajectories in the phase space, and allows 
to resort to ensembles.

Let me now come to the evaluation of the transition rates $R_\pm$.
From the preceding discussion, ${\cal F}$ determines the update frequency. 
At each trial, the probability that grains rearrange depends on the 
distortion of the force network, measured by the ratio $\sigma/P$.
The rates are thus written,
$$
R_\pm = R_0 \sqrt{T} e^{\pm \mu {\sigma/P}}
\quad,
$$
with some constant $R_0$, 
and where $1/\mu$ is a measure of how distorted the force network must be to induce shearing.
There are other ways of understanding why those rates should be functions of the ratio $\sigma/P$:
due to the hard-sphere interaction, no typical forces is determined by some interaction potential,
and $P$ appears as the only scale of force network; it has also been shown that the distribution
of forces is of the form $\exp(- a_0 F_i/<F>)$, indicating that the average normal force 
is also a measure of force fluctuations.~\cite{radjai98a,radjai98b}

For similar reasons, the local creation/destruction of arrangements performed by the flow,
is measured by $\omega^0=\ddpl\sigma/P$. Indeed, creation and destruction rates are expected
to be proportional to the strain rate. At a given deformation, $\dpl$,
the modification of the contact network might depend on $\sigma/P$, 
but not on any absolute value of the forces. Finally, the expression given to $\omega^0$ 
can be understood as a ratio of the work $\sigma\ddpl$ produced by the mean flow 
over some infinitesimal work $P\d l_0$ necessary to break a contact.

Let me now write the equations of motion~(\ref{eqn:stz:ddpl}) and (\ref{eqn:stz:npm})
in a more suitable way.
Variables
$$
\Delta = {n_--n_+\over n_\infty}\quad,\quad
\Lambda = {n_++n_-\over n_\infty}\quad,\quad {\rm and,}\quad
\sigp = {\sigma\over P}
$$
are introduced, along with the rescaled parameters $n_\infty = {2 {\cal A}_c / {\cal A}_a}$,
$\epsilon_0 = {{\cal A}_0\,{\cal A}_c/{\cal A}_a}$,
$\gamma = {\cal A}_0\,{\cal A}_c$, and $E_0 = 2\epsilon_0\,R_0$.
From~(\ref{eqn:stz:ddpl}) and (\ref{eqn:stz:npm}), it comes,
\begin{eqnarray}
\label{eqn:stzgr:1}
\ddpl&=&E_0\,\sqrt{T}\,
\left(
\Lambda\,\sinh\left(\mu\,\sigp\right)-\Delta\,\cosh\left(\mu\,\sigp\right)
\right)\\
\label{eqn:stzgr:2}
\dot\Delta&=&{1\over\epsilon_0}\,\left(\ddpl- \gamma\,\omega\,\Delta\right)\\
\label{eqn:stzgr:3}
\dot\Lambda&=&\gamma\,{\omega\over\epsilon_0}\,\left(1-\Lambda\right)
\end{eqnarray}
with $T$ and $\omega$ given by~(\ref{eqn:T}) and~(\ref{eqn:om}) respectively.
Variables $\Lambda$ and $\Delta$ represent respectively the total 
normalized density of STZ's and the bias between populations $n_\pm$.
These state variables account for an history-dependent texture 
(or fabric~\cite{radjai98a,radjai98b}) of the material,
and are one of the most interesting aspect of the current model.
However, since only steady state motions are considered in this work, 
$\Lambda$ can be safely taken to its asymptotic value, $\Lambda=1$. 
The fact that $\Lambda$ saturates to a constant value
means that the system adapts to the principal directions given by the stress tensor.
For the current problems, all the complexity of the dynamics lies in the values taken by $\Delta$.

Before attacking the case of a flowing layer of grains, let me consider 
a simple shear cell experiment, where the system is spatially uniform, so that 
$\omega=\omega^0=\sigp\ddpl$.
The system admits multiple jammed solutions, $\ddpl=T=0$, for any value of $\Delta$,
as well as a flowing solution for $\Delta = 1/(\gamma\sigp)$. 
In the flowing regime, the dynamics of rearrangements
as defined by equations~(\ref{eqn:stzgr:1}) and~(\ref{eqn:stzgr:2}) is stable so long as
$\tanh(\mu\sigp)>\Delta$. 
This condition is equivalent to $\sigp>\tan\Phi$, where $\Phi$ is the solution of,
$$
\tanh(\mu\tan\Phi) = {1\over\gamma\tan\Phi}
\quad.
$$
It is a Coulomb criterion of yield with limit angle $\Phi$.
Moreover, in the flowing regime, the stationary value of the temperature verifies, 
\begin{equation}
\label{eqn:mf:T}
T={\sigma\,E_0\,K(\sigp) \over \alpha}
\end{equation}
with, $K(\sigp) = \sinh(\mu\sigp)-\cosh(\mu\sigp)/(\gamma\sigp)$, whence,
\begin{equation}
\label{eqn:mf:ddpl}
\ddpl = \left(E_0 K(\sigp)\right)^{3/2}\,{\sqrt{\sigma}\over\sqrt{\alpha}}
\quad,
\end{equation}
which accounts for Bagnold's scaling.
 
Let me now consider a flow of granular material down an inclined plane.
Axis $x$ is taken along the descent and axis $y$ perpendicular to the plane.
Uniform solutions in direction $x$ are looked for, so that the variables $\ddpl$, $\Delta$,
and $T$ only depend on the height $y$. The velocity field is oriented 
along the direction $x$ and the velocity profile $u(y)$ is related to the strain rate by,
$u'(y) = 2\ddpl(y)$.
The total height of the layer of granular material is denoted $H$.
The plane makes an angle $\theta$ above the horizontal, 
and the stress tensor obeys Cauchy equations that enforce force balance,
leading to,
$P(y)= \rho g\,(H-y)\,\cos\theta$ and
$\sigma(y) = \rho g\,(H-y)\,\sin\theta$,
where $\rho$ is the mass density of the material, and $g$ the gravity.
The field $\sigp=\tan\theta$ is uniform.

Let me start with neglected spatial effects, $\omega=\omega^0$,
which corresponds to the limit $H>>\ell$.
The local equations are then identical to the mean-field case presented earlier.
For any $y$, the systems jams if $\theta<\Phi$, and flows otherwise.
In the flowing regime, the temperature and strain rate verify
equations~(\ref{eqn:mf:T}) and~(\ref{eqn:mf:ddpl}) respectively,
with $\sigma$ being a linear function of $y$.
Temperature $T$ grows linearly with the depth.
After integration over $y$, equation~(\ref{eqn:mf:ddpl})
leads to the following expression for the velocity field,
$$
u(y) = {4\over3}\left(E_0 K(\sigp)\right)^{3/2}\,{\sqrt{\rho g\sin\theta}\over\sqrt{\alpha}}
\left(H^{3/2}-(H-y)^{3/2}\right)
$$
The linear decrease of $T$ with $y$, 
this velocity profile as well as the rheology $u(H)\propto H^{3/2}$ agree
remarkably well with experimental and numerical findings.~\cite{pouliquen99a,silbert01}

\begin{figure}
\narrowtext
\begin{center}
\unitlength = 0.0011\textwidth
\begin{picture}(200,200)(-10,0)
\put(0,0){\makebox(200,200){\epsfxsize=180\unitlength\epsffile{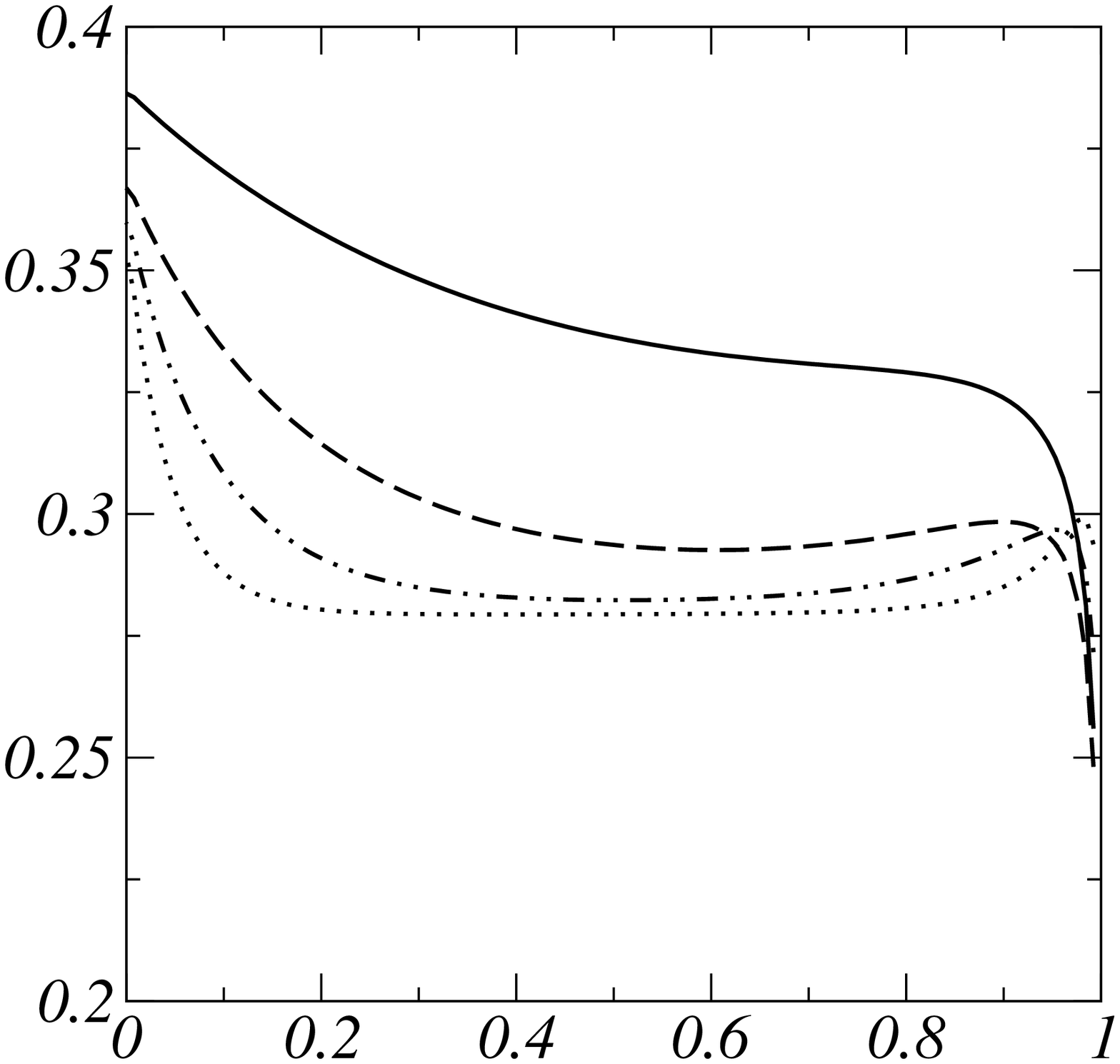}}}
\put(100,205){\makebox(0,0){(a)}}
\put(0,165){\makebox(0,0){$\Delta$}}
\put(170,0){\makebox(0,0){$y/H$}}
\end{picture}
\hspace{10\unitlength}
\begin{picture}(200,200)(-10,0)
\put(0,0){\makebox(200,200){\epsfxsize=180\unitlength\epsffile{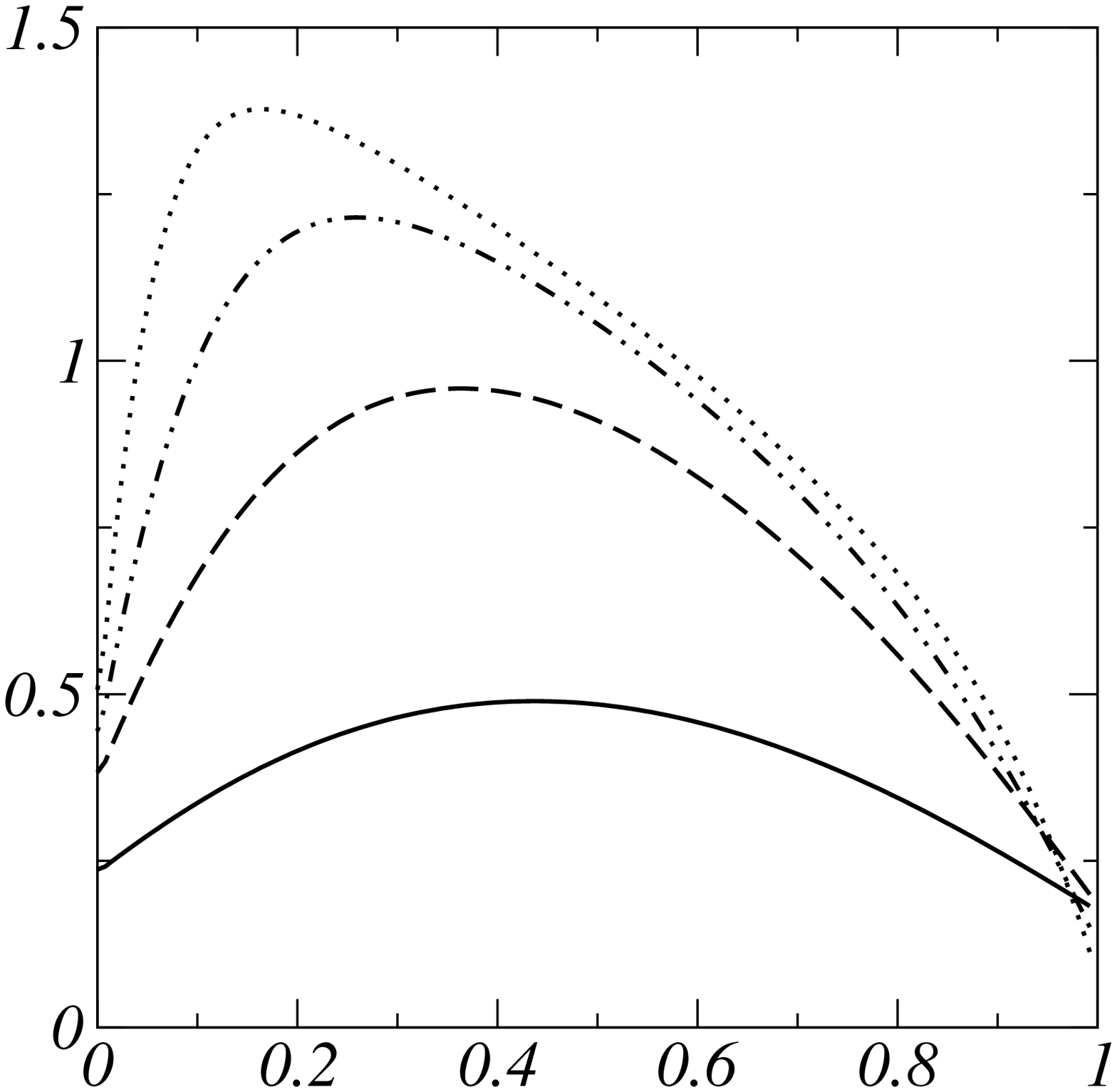}}}
\put(100,205){\makebox(0,0){(b)}}
\put(0,165){\makebox(0,0){$T$}}
\put(170,0){\makebox(0,0){$y/H$}}
\end{picture}
\vspace{20\unitlength}

\unitlength = 0.0011\textwidth
\begin{picture}(200,200)(-10,0)
\put(0,0){\makebox(200,200){\epsfxsize=180\unitlength\epsffile{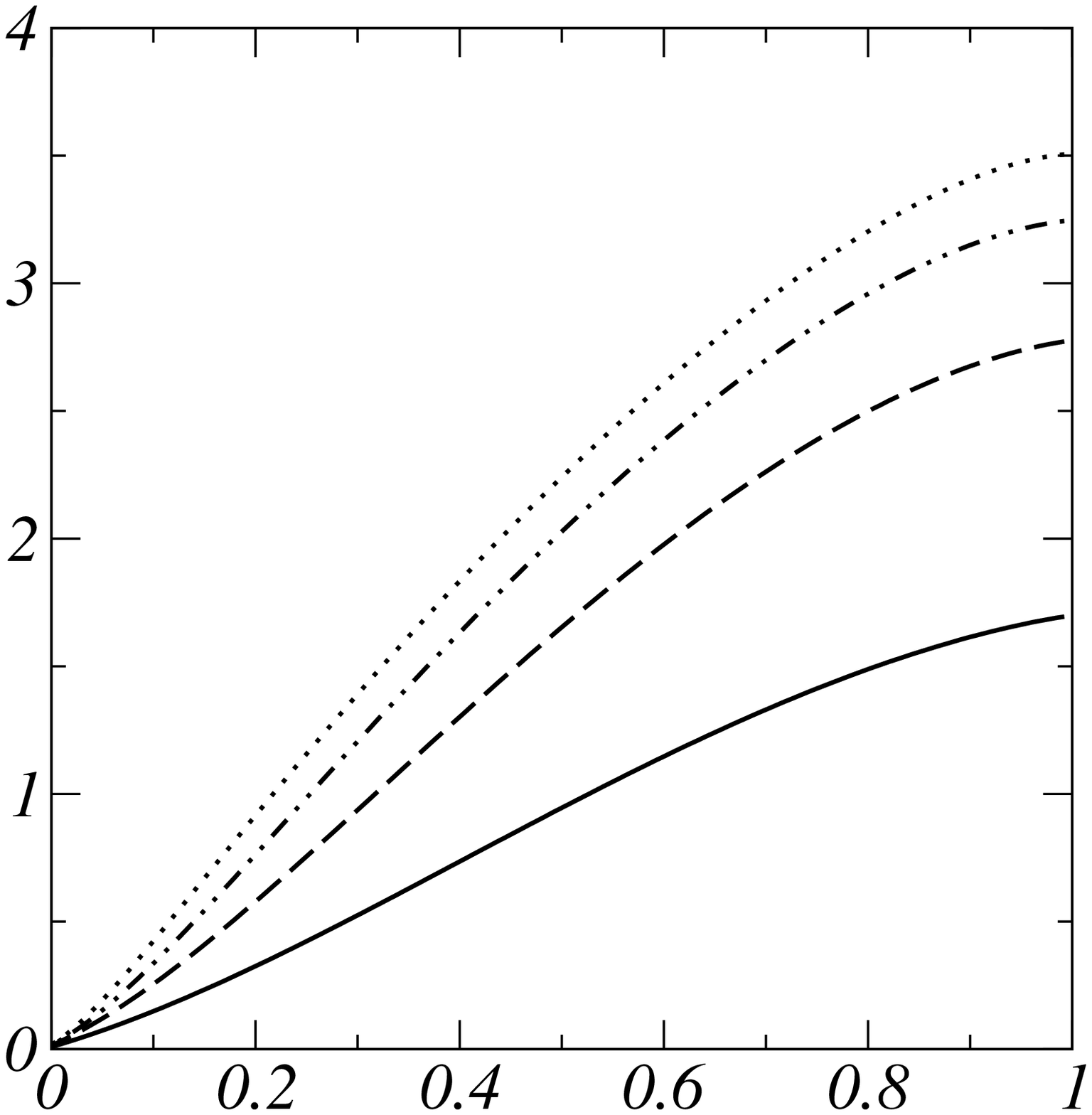}}}
\put(100,205){\makebox(0,0){(c)}}
\put(0,165){\makebox(0,0){$u$}}
\put(170,0){\makebox(0,0){$y/H$}}
\end{picture}
\hspace{10\unitlength}
\begin{picture}(200,200)(-10,0)
\put(0,0){\makebox(200,200){\epsfxsize=180\unitlength\epsffile{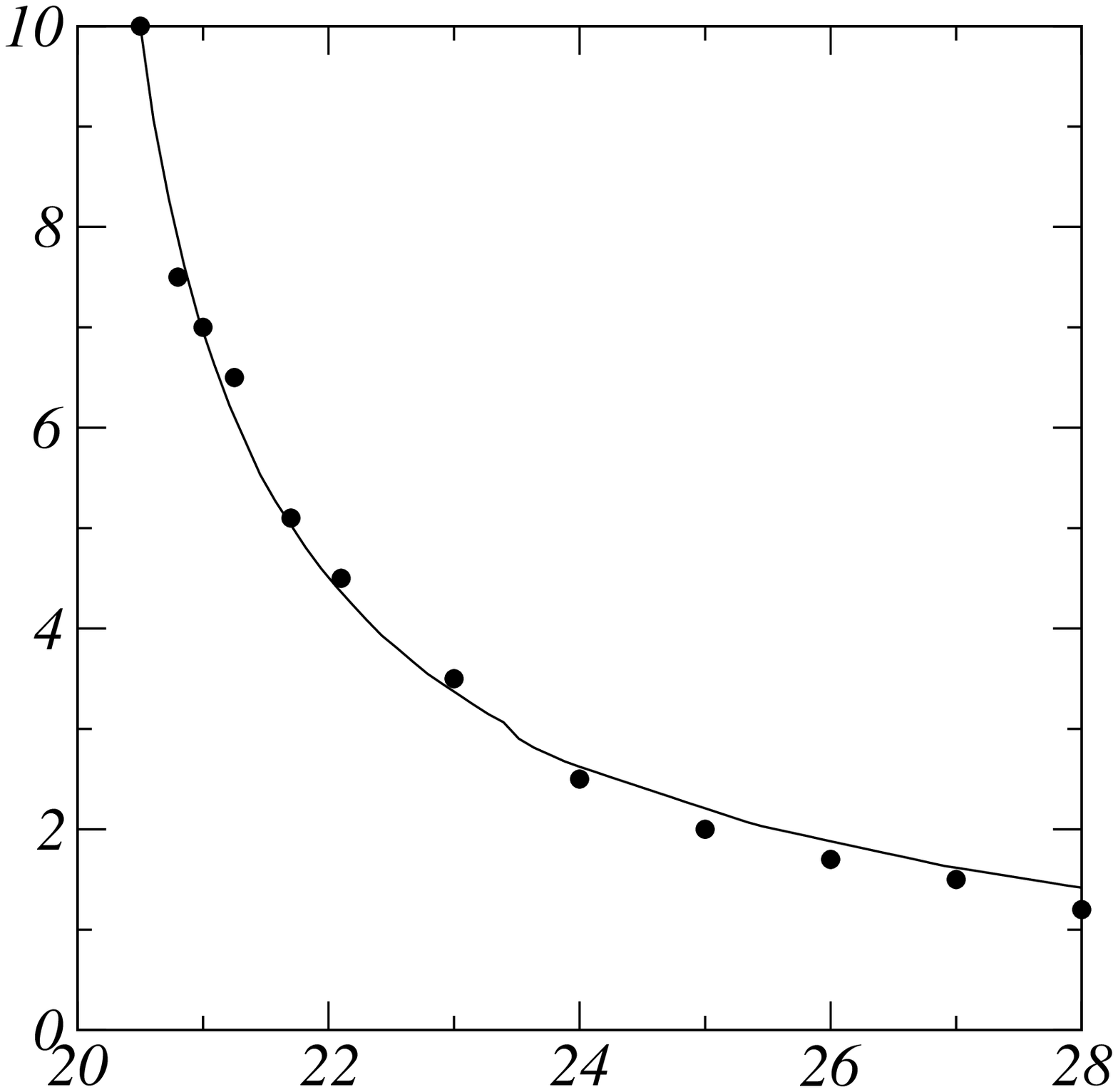}}}
\put(100,205){\makebox(0,0){(d)}}
\put(0,165){\makebox(0,0){$H_{\rm stop}$}}
\put(170,0){\makebox(0,0){$\theta$}}
\end{picture}
\vspace{20\unitlength}
\end{center}
\caption{
(a) The state variable $\Delta$ as a function of $y/H$, for values of the parameters 
taken from (d),
angle $\theta=24\deg$ and several values of $H$: 
$H=5$ straight line, $H=10$ dashed, $H=20$ dot-dashed, $H=40$ dotted.
(b-c) same for the temperature $T$, and the velocity profile $u$.
(d) $H_{\rm stop}$ from experimental data by Pouliquen~\protect\cite{pouliquen99a} (circles)
compared to the phase diagram obtained from the current theory (straight line)
with, $\gamma=10.23$, $\mu=0.98$, $\ell=0.99$. This curve does not depends on parameters
$E_0$ and $\alpha$.
}
\label{fig}
\end{figure}

In order to study the effect of the spatial coupling introduced {\it via} $\omega$,
it becomes necessary to resort to numerical integration.
The value of $\Delta$ in a steady state is now, $\Delta = {\ddpl/\gamma\omega}$,
and varies in space.
Typical profiles are displayed figure~\ref{fig}-(a).
The main effects of spatial interaction appear at the boundaries, resulting in an increase
of $\Delta$ close to the bottom, while $\Delta$ vanishes at the very top
(after a small bump due to the convexity of $\sqrt{H-y}$ near $y=H$).
The increase of $\Delta$ close to the bottom is due to the fact that the creation/destruction
of STZ result only from the action of the flow in the upper half plane $y>0$.
At the top of the flow, $\Delta$ vanishes because $\ddpl\to0$, whereas $\omega$ remains non-zero:
the system is more liquid-like, while it's more textured, solid-like at the bottom.

The non-uniformity of $\Delta$ has dramatic consequences on the stability of the flow.
The jamming criterion $\tanh(\mu\sigp)>\Delta$ is controlled by the maximum value of $\Delta$,
reached at the bottom of the flow.
Moreover, $\Delta$ globally increases with $\ell/H$:
if the lowest layers jam, the remaining flow of the upper layers presents
an even smaller effective height, the criterion for jamming is verified again,
and jamming propagates upward, resulting in a complete arrest of the flow.
This process leads to the existence of a critical height $H_{\rm stop}(\theta)$.
Such a curve is displayed figure~\ref{fig}
and compared with experimental data. The fit is remarkable.


The results of this work are two-fold.
Firstly, the role of activated rearrangements in granular matter has been evidenced,
hence providing a micro structural interpretation for the Coulomb criterion.
Secondly, the importance of granular temperature as opposed to thermodynamics temperature
has been shown; the fact that microscopic collisions determine the time-scale of
activated rearrangements lies at the root of Bagnold's scaling.
Those features contribute to the existence of a jamming criterion $H_{\rm stop}(\theta)$
and to temperature and velocity profiles
consistent with recent numerical results.~\cite{silbert01}

In any case, the remarkable agreement reached with various 
unexplained features of granular flows,
opens up new directions in our understanding of dense materials.

\acknowledgements
This work was supported by the W. M. Keck Foundation, and the NSF Grant No. DMR-9813752,
and EPRI/DoD through the Program on Interactive Complex Networks.

\end{multicols}

\end{document}